\documentclass[a4paper,twoside]{article}

\usepackage{epsfig}
\usepackage{subcaption}
\usepackage{calc}
\usepackage{amssymb}
\usepackage{amstext}
\usepackage{amsmath}
\usepackage{amsthm}
\usepackage{multicol}
\usepackage{pslatex}
\usepackage{apalike}
\usepackage[bottom]{footmisc}
\usepackage[english]{babel}
\usepackage{amsmath}
\usepackage{graphicx}
\usepackage{booktabs}
\usepackage[colorlinks=true, allcolors=blue]{hyperref}
\usepackage{SCITEPRESS}     

\begin{document}

\title{Tendencies in Database learning for undergraduate students. Learning in-depth or Getting the work done?}

\author{\authorname{Emilia-Loredana Pop\sup{1}\orcidAuthor{0000-0002-4737-4080}, Manuela-Andreea Petrescu\sup{1}\orcidAuthor{0000-0002-9537-1466},}
\affiliation{\sup{1}Department of Computer Science, Babes Bolyai University, Cluj-Napoca, Romania} 
\email{emilia.pop@ubbcluj.ro, manuela.petrescu@ubbcluj.ro} }

\keywords{Database, SQL, Computer Science, Student, Expectation, Challenge, Career Path, Basic Knowledge, Learning In-Depth, Getting The Work Done.}

\abstract{This study explores and analyzes the learning tendencies of second-year students enrolled in different lines of study related to the Databases course. There were 79 answers collected from 191 enrolled students that were analyzed and interpreted using thematic analysis. The participants in the study provided two sets of answers, anonymously collected (at the beginning and at the end of the course), thus allowing us to have clear data regarding their interests and to find out their tendencies.  We looked into their expectations and if they were met; we concluded that the students want to learn only database basics. Their main challenges were related to the course homework. We combined the information and the answers related to 1) other database-related topics that they would like to learn, 2) how they plan to use the acquired information, and 3) overall interest in learning other database-related topics. The conclusion was that students prefer learning only the basic information that could help them achieve their goals: creating an application or using it at work. For these students, ''Getting the work done'' is preferred to ''Learning in-depth''.}

\onecolumn \maketitle \normalsize \setcounter{footnote}{0} \vfill

\section{{\uppercase{Introduction}}}

Databases have become very important over the last decade, with an increasing impact on human's life, in almost all all fields of activities. Databases (DB, for short) mean various manners to save, access, modify and process data information, in or without combination with different applications/software, for the companies, public and private sectors of activity, education, students/people career path and personal life. 

In education, even if it is about prestigious universities, or the other ones (like, Anadolu University from Turkey \cite{Kamisli04}, Ghana \cite{Kwami15}, 
the Databases have a significant place in student management and in the knowledge process. The online databases allow access to documentation for work (see, for example, \cite{Kumar21}), open usage of the online learning journals 
and many other useful information, available in real-time. 


The high impact of Databases in our activities requires a specialized and dedicated human factor, such that this subject connected to SQL / NoSQL became mandatory for the students of Mathematical and Computer Science, Technical and Engineering specializations. Due to \cite{Kahraman22,Spieler20}, the student's socioeconomic status, their motivation, performance and self-efficacy, the task value beliefs and the engagement in Computer Science, are also important aspects that influence the knowledge and involvement to obtain performance in Databases.  

Nowadays, in the labour market, due to the high number of students enrolled in Computer Science specializations, start seeing changes related to how the learning with passion and involvement was transformed in {\em search on the Internet} or {\em just want to create the application}. The modifications became more and more visible, like, {\em the learning in-depth} trend was transformed in {\em get the work done} trend. 
Because Databases are of actual interest, our research, which involves it, could be addressed to the entire community, giving different perspectives and alternatives from the student's points of view, our new generation, that will lead us to the future, being the new employees and employers. 

Our analysis related to Databases involves the students enrolled in the second year in different specializations and lines of study. Our faculty, Faculty of Mathematics and Computer Science, from Babe\c{s}-Bolyai University, Cluj-Napoca, Romania, is directly connected to the IT domain of activity and it is specialized in learning the students, the basics of IT, like programming languages and Databases. We have applied a survey to our Databases course, in which the participation was optional and anonymous (we only asked them to choose their specialization and their gender), to increase the freedom of the student's responses.  
The scope of the paper is to present the student's expectancy related to Databases and SQL-related domains and how could be correlated with the requests from the labour market. We analyzed their interests and knowledge about Databases/SQL at the beginning of the course (some of them, had previous experience, but others did not know the notions) and also at the end of it (we performed a final survey in the last week of the course). Then, we also wanted to find out if the students prefer a dedicated career path in Databases/SQL, database security, or efficiency, or they just want to make use of it seldom. 
In order to obtain the desired answers, we have asked the students to complete the following open questions: {\em R1: Do you have other DB-related topics that you would like to learn about, besides the ones from the Databases course?}; 
{\em R2. Did your expectations related to the Databases course been covered? How do you plan to use the learned information?}; {\em R3. Are you interested in learning other Database-related topics, or working in DB-related fields (as Database Administrator, SQL Data Analyst, or Database Developer)?}.

The paper has the following structure: it starts with an {\em Introduction} and a {\em Literature Review}, in which are presented articles and papers on the same topic subject, followed by the {\em Methodology} section, the {\em Threats to Validity} section, that include the possible threads of validity and the actions performed in order to minimize and mitigate them, and it ends with the {\em Conclusion and Future Work} section, in which is summarized the work with the obtained results and also are mentioned some future approaches. The most important section, {\em Methodology}, presents the methods used for this survey to obtain the results, starting with the participant's set, course curricula, asked questions, and continuing with the data analysis of the received responses, 
and then, for each of the research questions, were provided relevant conclusions, like, the student's expectations in the Databases course proved to be satisfied and their achieved knowledge being focused only on the basics (for most of the students), homework proved to be one of the challenges (for most of the students), a career path of Databases/SQL seemed to be uninteresting, and the most important aspect, {\em learning concepts in-depth} trend transformed in {\em get the work done} trend.  



\section{{\uppercase{Literature Review}}}

During the last decades, Databases came closer to people's careers, by bringing the possibility of storing and accessing, in different perspectives and alternatives, big amounts of data easier, more efficiently, and in an useful manner.  
Connected with the Computer Science field of activity, Databases are represented in the industry with career positions like: Database Administrator, Database Migration Operator, Relational / Non-Relational / SQL / NoSQL  Database Developer, Database Manager / Architect / Engineer, Data Operator, Data Scientist / Data Analyst,  Business Analysts / Business Intelligence Developer, and many other more (see, for example, \cite{Jaiswal22}). 

In the process of choosing the desired career path a very important aspect has to be given to efficiency and also should be taken into account the comparisons between the genders involved \cite{Mann20}. 
For the students from the Technical University of Iasi, Romania, have been identified their career aspirations, as: lifelong learning, desire of continuing their studies after finishing the actual ones, and participation in different training courses \cite{Anghel15}.
For the undergraduate IT students of an Australian university, were analyzed the career aspirations, by using an online self-assessment of study and career confidence referring the discipline and of a survey about short-term and longer-term career aspirations and prior experience \cite{Mckenzie22}, revealing the best motivation as being the intrinsic interest and enjoyment of IT, without considering the time involved in such a position. 
Another analysis, based on questions and involving a variety of disciplines, was performed, in Australia, for the students (undergraduate, postgraduate) of a representative university, and concluded that, before graduation students proved to be uncertain about their careers, but after graduation, the career path became clear and quite easy to be identified \cite{Kinash17}. 


For the Databases field of activity could be identified studies that contain relevant aspects and arguments, some of them obtained from applied surveys to the students, professors, and various members of the institutions involved \cite{Dehghani18,Uzun20}. 
The educational methods and their limitations were analyzed, for postsecondary education, in a survey of the published articles from 2015, a process performed by the United States' National Center for Education Statistics \cite{Mendoza17}.  
Hybrid learning (online and onsite) was proved to be the optimum learning method for sixty-two college students in the U.S., after an online survey was applied, one year after the outbreak of COVID-19 \cite{Zhou21}.
Databases proved to generate a high degree of satisfaction in the activities performed by the students and by the faculty members and also in their lives, at the Shahid Beheshti University, where a survey with a descriptive approach (questionnaire and log files) was proposed \cite{Dehghani18}.  
At the Department of Computer Education and Instructional Technology, Faculty of Education, Uludag University, Turkey, was applied a semi-structured interview in which participated 25 students from the second year, having as final result the analysis of the problem-based learning method applied to the course Database Management Systems, that revealed the importance of a basic level of knowledge in the implementation of the problem-based method for designing the classroom environment \cite{Uzun20}. 

Connected to the Databases field of activity, in Computer Science and Software Engineering, Structured Query Language (SQL) skills are mandatory. SQL represents the most important element of Databases that provide relevant data information for a potential user of any database and also it is used in education, being the basic language of Databases.  
Besides the knowledge of SQL, in education is also required pedagogical skills. 
Some teaching practices in SQL education and also a systematic map of educational SQL research with a future research agenda are presented in \cite{Taipalus20}, with recommendations for the educational SQL research to include studies on advanced SQL concepts and on aspects not related to data retrieval and also replication studies. 
For some people, Databases mean SQL, NoSQL, and Big Data, but for others can mean job positions, like, Data Analyst, Data Science, and Business Intelligence. What's for sure, is the fact that Databases store big amounts of data information in simple or complex and different ways and assure access to them. Both SQL and NoSQL Databases are in search nowadays to complete the business trends and patterns, and near to them are the Data Warehouses used for the technological skills and strategical and statistical competencies. An analysis related to the presentation of the coordinates used in processing data and implications for the academic curricula that also provide arguments for Data Analyst and Business Intelligence job positions in the idea of acquiring a corresponding level of SQL and of Data Warehouses knowledge is presented in \cite{Fotache15}. 
To learn Databases, SQL / NoSQL is easy, because everywhere the information is available: online training, documents/courses (for example, \cite{Halvorsen17,IBM10}), articles, videos, and their presentations contain and present in various manners the basic and complex notions involved. 


\section{{\uppercase{Methodology}}} 

\textit{How}: 
According to ACM Sigsoft Empirical Standards for Software Engineering Research \cite{ACM}, our method is classified as survey research. We used surveys that contained open and closed questions, closed questions were used to assure participant representations related to the study line or to measure their intentions and the open questions were used to obtain relevant and in-depth information. The students were asked to answer closed questions that could allow us to measure their intentions, for example {\em To which extend do you plan to use the acquired knowledge?} having a scale from 1 to 5 (\textit{Definitely NOT Use} to \textit{Definitely Use}), but also open questions: \textit{How do you plan to use the acquired knowledge?} to have a better understanding of their intentions. We wanted to see how students' perception changes over time, so we had two surveys, at the beginning and during the last week of the course. We were interested in the student's challenges related to the course, how their perception of following a career in database-related fields after taking the course evolved, and to what extent they plan to use the acquired knowledge. 
To find out we asked the students enrolled in the Databases course from different lines of study (Computer Science - English, Computer Science - native language, and Mathematics and Computer Science - native language); as the course syllabus was identical for all the lines of study. The course topics (or, course syllabus) are the ones given in Table \ref{tab:curricula}.
\begin{table}
  \caption{\textit{Fundamentals of Databases} syllabus}
  \label{tab:curricula}
  {\small{
  \begin{tabular}{p{7.0cm}}
\toprule
Lecture 1. Introduction to DBs. Fundamental Concepts. \\
Lecture 2. The Relational Model. \\
Lecture 3. SQL Queries.\\
Lecture 4. Functional Dependencies. \\
Lecture 5. Normal Forms. \\
Lecture 6. Relational Algebra.\\
Lecture 7. The Physical Structure of DBs. \\ 
Lecture 8. Indexes. \\
Lecture 9. Binary trees. ISAM. 2-3 trees. B-tree. B+ tree. \\
Lecture 10. Object Oriented Databases. \\
Lecture 11. Query Optimization in Relational Databases. Evaluating Relational Algebra Operators. \\
Lecture 12. Transactions. Concurrency Control. \\
Lecture 13. Conceptual Modeling. Data Streams. \\
Lecture 14. Short review: Relational DBs. Appendixes. \\
\toprule
\end{tabular} 
}}
\end{table}

\textit{Research Ethics}: The survey was optional and anonymous, and the students were informed about these aspects and also about the purpose of collecting and using the information they provided (there were no other purposes than the mentioned ones). 
We had 79 answers from 191 students enrolled in the course.

\subsection{Participants}

Being an optional survey, only 79 students from 191 enrolled students participated in the survey and provided answers, 42\% of all the students. The participants were students enrolled in the first semester of the second year of study. The survey was open for two weeks every time, thus allowing students to submit their responses, after two weeks we closed the survey, as we considered that the students who did not answer, probably don't want to participate in the study, so no other answers will be collected. We considered that in terms of percentage and numbers, the number of answers is comparable with the number of answers from other studies from the Computer Science domain \cite{marwan2020,enase21,icsoft22}, thus making this study a valid one from this point of view.

\subsection{Data Collection and Analysis}
\label{sec:data_collection}

The responses were collected using Google forms, we send the survey link to the students, who completed it, at their own pace (even if completing the survey took approximately 5 minutes). We saw an advantage in the short answers provided by the students, sometimes the answers contained typing mistakes because we considered their answers to be genuine, without too much thinking and analyzing, thus providing more accurate data. The answers were collected anonymously, and most of them were submitted on the same day when they were asked to participate in the study.
We used quantitative methods (questionnaire survey according to ACM community standards specifications \cite{ACM}) to obtain the data and for interpreting the text we used thematic analysis \cite{Braun19}. Other Computer Science related studies used the questionnaire survey method 
{\cite{redmond13,easeai22}} and thematic analysis \cite{Kiger20,enase21}. According to the methods mentioned, we organized this study in two steps: the first one was to obtain the data (taking into consideration research ethics) and the second step was to analyze it: 
\begin{itemize}
\item 1. Analyzed the answers and restructured them if necessary. Sometimes parts of the answers/whole answers were more appropriate as answers to other questions.
\item 2. Found specific keywords and grouped them into classes (tasks performed by one author).
\item 3. The classification was analyzed by the other author, and both authors discussed if changes are needed. 
\end{itemize}

Some answers were short (with just one or two words), others were more descriptive, and we identified more than one or two keywords in a specific answer. Due to this aspect, the total number of specific keywords mentioned is greater than the number of students participating in the study. 
We asked the same closed questions in both surveys, to aggregate the groups (Qg1, Qg2), but the open questions were different, as we wanted to analyze how student's perceptions changed, what were the major course challenges and how they envision themselves as working with knowledge acquired in this course (see, Table \ref{tab:curricula1}).
 
\begin{table}
  \caption{\textit{Fundamentals of Databases} syllabus}
  \label{tab:curricula1}
  {\small{
  \begin{tabular}{p{7.0cm}}
\toprule
    Qg1. Line of study (choice of \textit{Computer Science / Mathematics and Computer Science} \\
    Qg2. Gender (Choice of \textit{Male / Female / I don't want to answer})\\
\midrule
    \textit{Quiz 1 questions:} \\
    Q1. What are your expectations related to the Database course? \\
    Q2. Do you have DB-related knowledge? Do you know what is an SQL statement? \\
    Q3. How do you plan to use the learned information? Are you interested to work in DB-related fields (for example as Database Administrator, SQL Data Analyst, or Database Developer)? \\
\midrule
    \textit{Quiz 2 questions:} \\
    F3. How did the Databases course meet your expectations? (closed answers using a five points scale from: under expectations to exceeded expectations)\\
    F4. What challenges did you encounter during the Databases course? \\
    F5. What other topics related to Databases are you interested in (and were not studied in the course)?  \\
    F6. How do you plan to use the Databases course knowledge?\\
    F7. Do you see yourself working in the Databases (as the main field of activity)?\\
\toprule
\end{tabular}
}}
\end{table}

\subsection{Q1: What were your expectations related to the Databases course and were the expectations met?}

\begin{figure}[htbp!]
    \includegraphics[width=0.5\textwidth]{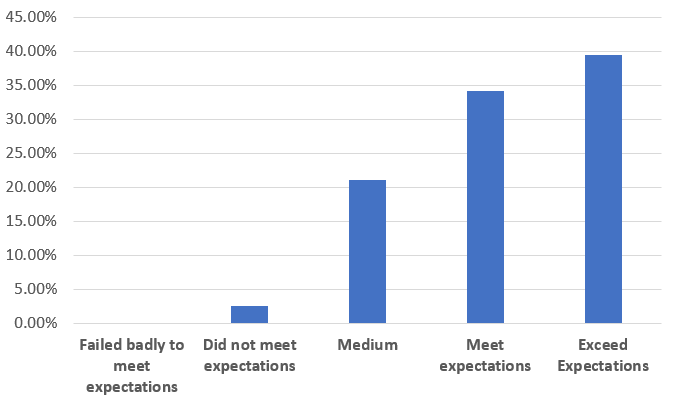}
    \caption{Course Expectations}
    \label{fig:meet}
\end{figure}

The Databases course was an introductory one, so basic notions were introduced. The course was delivered with the assumption that students did not work or learn Database notions, but the reality was a little bit different as there were students that previously worked with SQL and databases: \textit{"I had an internship as a software engineer where I worked with SQL", "I currently use some of the principles at work"}. However, most of the students did not work with database and/or SQL queries\textit{"My database knowledge is practically nonexistent", or "I have NoSQL or Database knowledge"}. Due to this aspect, 84\% of them stated that they want to learn databases basics. 
Analyzing the answers, we found out that students don't want to learn in-depth, to learn about complexity, security, or efficiency, their expectations are related to acquiring knowledge that would allow them to use SQL and databases. Only 31\% mentioned that they would like to learn about database administration, as most of them wanted to learn SQL: 58\% and 84\% basic concepts. 
At the end of the course, we wanted to find out if their expectations were met, the question was a closed one, allowing us to measure their perception; the results can be seen in Figure \ref{fig:meet}.

In conclusion, students' expectations were related mainly to achieving basic knowledge in the database-related field, and SQL instructions; The course meet their expectations.

\subsection{Q2: What were the course's main challenges?}
\begin{figure}[htbp!]
    \includegraphics[width=0.5\textwidth]{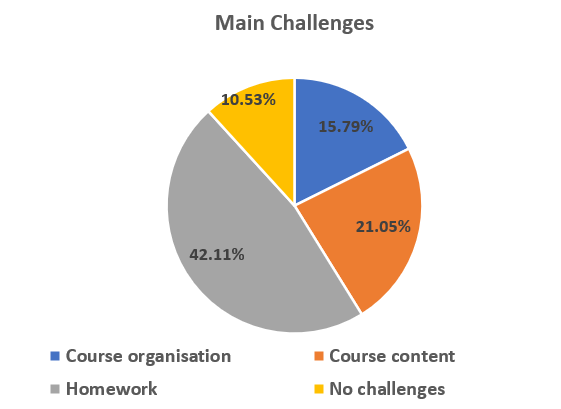}
    \caption{Course Challenges}
    \label{fig:challenges}
\end{figure}

At the end of the course, we wanted to find out what challenges were encountered by the students, taking into consideration that the course involved lectures, seminars, laboratory work, and homework. Each homework assignment was graded and the average grade was a part of the final course grade, so the students had an interest in solving their homework correctly and on time. By analyzing their answers, we found out that their challenges can be aggregated into four main groups related to course organization - 15.79\%, course content - 21.05\%, course applicability /homework - 42.11\%, and some students reported that they did not encounter any challenges - 10.53\%. The results can be seen in Figure \ref{fig:challenges}.

The students mentioned organizational issues: \textit{''the seminars are not posted on time, before the lab. We need the information during the lab, so it would be nice to have access to it before the lab.'', ''there are a lot of communication channels: team channels and the site''}. In terms of course content, the major complaints were related to triggers and procedures, or even to definitions: \textit{''triggers were not easy to understand''} and \textit{''procedures were a headache to go through and understand, but other than that the course support materials were really helpful and descriptive'', ''The definitions which were with greek letters, those were very complicated to understand''}. The major challenges reported were related to homework, some students mentioned the text was not as clear as expected: \textit{''Some homework text was a bit obscure''} or the task was not trivial: \textit{''some parts of the homework were somehow difficult'', ''Slight difficulty with some of the subpoints from the last lab''}.
In conclusion, most of the students encountered challenges related to homework, and some of the observations can be implemented (such as defining more clear assignments). As for the homework difficulty, we consider that it's beneficial for a student to have ''slightly'' difficult homework as solving the problems is a step in acquiring knowledge.

\subsection{Getting the work done or learning concepts in-depth?}

To find out the answer to this inquiry, we asked two questions and correlated the answers: \textit{''What other topics related to Databases are you interested in (and were not studied in the course)?''} and \textit{''How do you plan to use the Databases course knowledge?''}. The students understood the importance of being able to work with databases: \textit{''things that could help me in the future'', ''to create a database having a good table structure''} or \textit{''I'll need this information when I will develop my applications/sites''}, even if some aimed to learn only the most basic information: \textit{''SQL statements are not important as I can find them on internet. We should know only that they exist, how are called and used for, and then we could search for them on the internet''}.
To the question \textit{''How do you plan to use the Databases course knowledge?''}, we group their answers into five categories: \textit{''Definitely NOT use database knowledge'', ''Probably NOT use'', ''Maybe'', ''Probably use''} and \textit{''Definitely NOT use''}. The percentages can be seen in Figure \ref{fig:useIntent}: 
 \begin{figure}[htbp!]
    \includegraphics[width=0.5\textwidth]{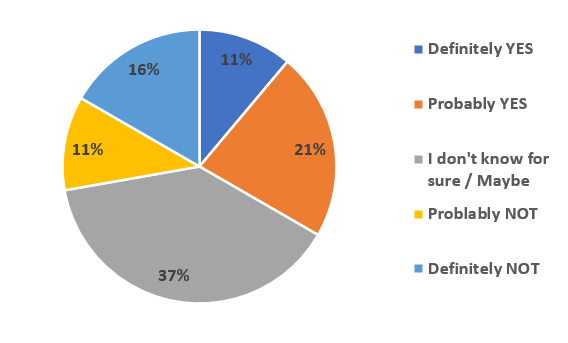}
    \caption{Use Intentions}
    \label{fig:useIntent}
\end{figure}

\begin{figure}[htbp!]
    \includegraphics[width=0.5\textwidth]{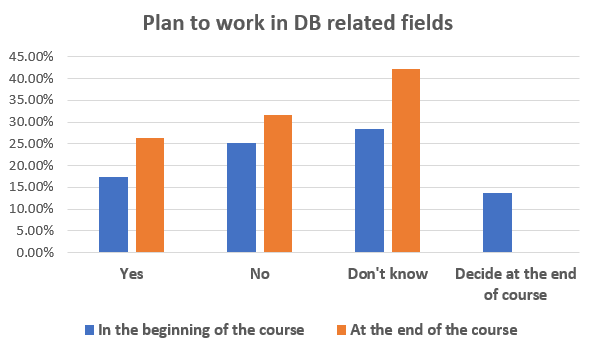}
    \caption{DB Related Work Interest}
    \label{fig:WorkPlan}
\end{figure}

Some students stated that they don't know how or when will they use database-related knowledge \textit{''I don't know''}, some stated that they plan to use them \textit{''only tangential when I have to'', ''when I have to develop an application''}, and most of the students that mentioned how will they use it was for \textit{''personal projects''} and \textit{''at work''}. A few answers stated that \textit{''I already use them a lot at work''}. The low interest in learning in-depth is correlated to their low interest in following a career path in database-related fields.  However, in total, the percentage of students that would not choose to work in a DB-related field was 25\% in the first quiz compared to 31.58\% in the second quiz (end of course). The percentage of students that want to work in a database-related field increased from 17.43\% in the beginning to 26,32\%. The number of students that did not decide increased from 28.5\% at the beginning of the course to 42.11\% at the end of the course, probably the group of students that planned to decide at the end of the course, did not take any decision. The percentages and evolution are reflected in Figure \ref{fig:WorkPlan}.

Based on these values, we can assume that students are not interested to follow a career in DB-related fields, and consequently, they do not want to learn in-depth database mechanisms.
To the question \textit{''What other topics related to Databases are you interested in (and were not studied in the course)?''}, we group their answers into seven categories: \textit{''Practical Examples'', ''NoSQL'', ''Deployment'',	''Efficiency'',	''Security'', ''Request Based'', ''Nothing''}. Some answers contained more than a keyword, so the summarised keywords' percentage appearance is greater than 100\%. The percentages can be seen in Figure \ref{fig:examples}:

 \begin{figure}[htbp!]
    \includegraphics[width=0.5\textwidth]{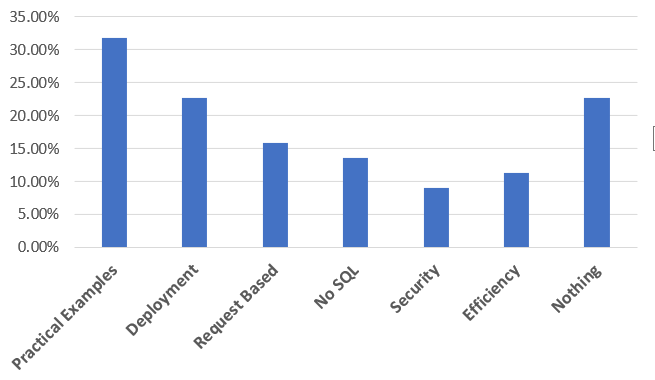}
    \caption{Knowledge Interest}
    \label{fig:examples}
\end{figure}

Students' answers come in all lengths, from the shortest ones: \textit{''NoSQL'', ''Cloud Database''} to a maximum of 183 words, but the average answer was about 16 words. Some stated that no other information should have been taught in the course \textit{''Nothing, it was perfect.''} to a demand for getting the work done \textit{''to create a database as simple and fast as possible'', ''database deployment''}, to a pledge for other topics \textit{''we should quit Microsoft SQL and learn something that can work on multiple platforms, maybe some non-relational stuff''}. However, a large part of the students (36,84\%) 
mentioned they wanted more practical examples, asking for a solution to a particular situation: \textit{''We should do more practical examples related to BLOB types and how are they used in a database''} or \textit{''I was expecting to connect a database to a known language (Java for example)''}.
As a conclusion for the last question, the students were not that interested in learning new concepts (the students' percentage that mentions new technologies and concepts is significantly lower compared to the percentage of students that wanted to get ''the work done'').  We considered that ''get work done'' includes the request for practical examples (31.82\%), the request for deployment (22.37\%), and request-based percentage (based on the needs of the current project 15.91\%). The mentioned aspects appear as keywords in 70.45\% of the answers compared to the summarized percentage for \textit{''No SQL'', ''Security''} and \textit{''
Efficiency''} which is 34.09\%. 
We analyzed the results and we took into consideration the following major arguments related to the database introductory course:\\
\textbf{Argument A. Students are interested in learning only the basics.} A large majority of students 84\% mentioned that they want to learn and know the basics, but the interest in learning (complexity, security, etc.) was minimal, given by 13,46\% of the students.\\
\textbf{Argument B. Students' challenges regarding were mainly related to homework/practical part}. The information presented at the course represented a challenge only for a small part of them, no one mentioned reading or having challenges with implementing more than was requested by the teacher. The students were interested to do their homework but at the bare minimum.\\
\textbf{Argument C. Students want to learn only \textit{''what is needed to create my app''}}. We analyzed the answers received regarding how they plan to use the information, most of them stated practical examples and wanted to know just enough to create and work with a database in an application, or when needed \textit{''at work''}. They just want ''to get the work done''. \\
\textbf{Argument D. What do they want to learn next?}. The students' percentage satisfied with the acquired information level is doubled compared to the percentage of students who want to learn advanced features/ other topics related to databases. In conclusion, a large part of them in not interested in learning more database-related topics.

Based on the previous arguments, we concluded that most of the students prefer to ''get the work done'' without learning or making too much effort in acquiring new knowledge and skills, as they expect to learn \textit{"without stress"} because \textit{''we can search for information on Google''}. \textbf{Learning in-depth} or \textbf{Getting the work done}? For these students, the answer is ''Getting the work done''.

\section{{\uppercase{Threats to Validity}}} 

We focused on mitigating the threats to validity as they were defined in \cite{ACM} and we analyzed the following topics: target participant set, participants selection, contingency actions for drop-outs, and also research ethics. 
The target participant of our survey was represented by a set of students, from the second year, enrolled in the Databases courses, on the specializations {\em Computer Science} and {\em Mathematics and Computer Science}, in the Romanian and English lines of study. 
The students that participated were organized in groups of study, alphabetically ordered by their surname. The groups were randomly selected, so each participant could be part of the survey, without any other selection criteria. Because of the random selection of the student groups, no other threats of the participant set and participant selection were need it. 
As the involvement of the students in the survey was optional, the methods used by us were limited to enlarge the number of participants. Our best option was to explain clearly and deeply the survey purpose and its final results, which could significantly increase the quality of our activities in Databases (27 versus 52 responses).
As for research ethics, the students knew the purpose of the survey and also that their participation was optional (shown by the response rate). Moreover, the students were able to choose the questions from the survey that they wanted to answer. 
Another threat taken into consideration refers to our approach to processing the data, that we tried to make it less subjective, by following the recommendation for the data processing and also by checking others' work. 


\section{{\uppercase{Conclusion and Future Work}}} 


For the second-year students, enrolled in Computer Science and Mathematics and Computer Science specializations, we wanted to analyze their perception related to Databases, their importance, and how attractive could be a job position in a database-related domain. A survey was applied, to all the 191 students that participated in the course. We required information related to their expectations, knowledge level, and interest in working in a database-related field of activity, at the beginning of the course and also at the end of it. The survey was optional and anonymous and all the students were encouraged to participate in it. The analysis of the responses was performed as the community standards recommended, and all the possible threats to validity were handled, by having a diverse set of participants (given only by specialization and gender).
We concluded that most of the students are interested in a basic level of Database knowledge and they preferred to focus only on the SQL knowledge part. The challenges met during the semester, proved to be, for most of them, related to the homework. 
Most of the students, that were undecided if they want or not a job related to Databases/SQL (DB administrator, SQL developer, and so on), in the beginning of the course, concluded that are not interested in a database career path, in the end of the course. They just want to consider the job positions available in the labour market, even if are or are not connected to the database-related domain. They also consider the knowledge of Database basics as being enough for their career path.
Our main conclusion, of this survey, was identified, as being, the replacement of the {\em learning in-depth} concept with the {\em get the work done} concept, a principle applied by the students during the semester.  
In the future, we want to find out if the same trends and aspects could be found in other courses, too, to provide comparisons related to the gender of the students involved and also to analyze a long-time perception.  \\

{\bf {\uppercase{Funding}}}
The publication of this article was supported by the 2022 Development Fund of the Babe\c{s}-Bolyai University.

\bibliographystyle{apalike}
{\small
\bibliography{example}}

\end{document}